\documentclass[aps,prb,twocolumn]{revtex4}
\usepackage{epsfig,amsmath,amssymb,color} 
\newcommand{\be}{\begin{equation}}
\newcommand{\ee}{\end{equation}}

\newcommand{\ra}{\rangle}
\newcommand{\la}{\langle}
\newcommand{\bit}{\begin{itemize}}
\newcommand{\eit}{\end{itemize}}
\newcommand{\bea}{\begin{eqnarray}}
\newcommand{\eea}{\end{eqnarray}}

\usepackage{epsfig}

\begin{document}
\title
{The spin-half Heisenberg antiferromagnet on two Archimedian lattices: \\
From the bounce lattice to the maple-leaf lattice and beyond }
\author
{D. J. J.~Farnell$^1$, R.~Darradi$^{2}$, R.~Schmidt$^{3}$, and J.~Richter$^{3}$}
\affiliation
{
$^{1}$Division of Mathematics and Statistics, 
Faculty of Advanced Technology, University of Glamorgan, 
Pontypridd CF37 1DL, Wales, United Kingdom\\
}
\affiliation
{
$^{2}$Institut f\"ur Theoretische Physik, Technische Universit\"at Braunschweig, D-38106 Braunschweig, Germany\\
}
\affiliation
{
$^{3}$Institut f\"ur Theoretische Physik, Otto-von-Guericke Universit\"at
Magdeburg, 39016 Magdeburg, Germany \\   
}

\begin{abstract}
We investigate the  ground state 
of the two-dimensional Heisenberg antiferromagnet 
on two Archimedean lattices, namely, the maple-leaf and bounce
lattices as well as a generalized  $J$-$J'$ model interpolating between both
systems by varying $J'/J$ from $J'/J=0$ (bounce limit) to $J'/J=1$
(maple-leaf limit) and beyond. We use the coupled cluster method to high orders of 
approximation and also exact diagonalization of finite-sized lattices 
to discuss the ground-state magnetic long-range order based on  
data for the ground-state energy,  the magnetic order 
parameter, the spin-spin correlation functions as well as the pitch angle
between neighboring spins.
Our results indicate that the ``pure'' 
bounce ($J'/J=0$) and maple-leaf ($J'/J=1$) Heisenberg 
antiferromagnets are magnetically ordered, however, with a sublattice
magnetization drastically reduced by frustration and quantum fluctuations.
We found that magnetic long-range order is present in a wide parameter range  
$0 \le J'/J \lesssim J'_c/J $ and that the magnetic order parameter 
varies only weakly with $J'/J$.
At $J'_c \approx 1.45 J$ a direct
first-order transition to a quantum orthogonal-dimer 
singlet ground state without magnetic long-range order takes place.
The orthogonal-dimer state is the exact ground state in this large-$J'$ regime, 
and so our model has similarities to the  
Shastry-Sutherland model. Finally, we use the exact diagonalization
to investigate the magnetization curve.
We a find a $1/3$ magnetization plateau for $J'/J \gtrsim 1.07$ and another one
at $2/3$ of saturation emerging   
only at  large
$J'/J \gtrsim 3$.
\end{abstract}
\pacs{Valid PACS appear here}
\maketitle


\section{Introduction}
The study of  two-dimensional (2D) quantum spin-half antiferromagnetism 
is an interesting and  challenging problem, in particular, if the magnetic
interactions are frustrated. \cite{Man,balents} 
In 2D systems the interplay between geometry and quantum fluctuations may
lead to semi-classical ground state (GS) phases with conventional magnetic
long-range order (LRO) as well as to new quantum  phases without magnetic
LRO. \cite{Lhu,wir04}  
The spin-$\frac{1}{2}$ Heisenberg antiferromagnet (HAF) on the eleven
2D Archimedian and related lattices presents an excellent opportunity to investigate
the subtle balance of interactions and fluctuations and the role of lattice
geometry.
\cite{wir04}     
It is well-established that magnetic LRO is present in the GS 
of the spin-$\frac{1}{2}$ HAF on bipartite lattices  (square \cite{Man,wir04}, 
honeycomb\cite{reger89,Mat,wir04}, 1/5-depleted square (or CaVO)
\cite{Tro,Ma,wir04}, 
square-hexagonal-dodecagonal\cite{ToRi,wir04}).
However, this magnetic order can be weakened or even suppressed
by the presence of frustration. 
The first investigation  of a frustrated quantum   
HAF goes back to the early 1970s, when Anderson and 
Fazekas\cite{And,Faz} considered the spin-$\frac{1}{2}$ HAF on the 
triangular lattice. They conjectured a magnetically disordered GS.
However, it is now clear that there is magnetic GS LRO in this
system, see, e.g., Refs. \onlinecite{Ber1,chub94,Cap,farnell01,wir04,farnell09}. 
The pioneering work of  Anderson and Fazekas has formed the starting point for an
intensive  
investigation of frustrated quantum magnetism.
In particular, it has stimulated the search for non-magnetic quantum states
in 2D magnetic systems.

Another well investigated frustrated 2D system is the HAF on the SrCuBO 
lattice, which can be transformed by an appropriate distortion to the 
Shastry-Sutherland square lattice model\cite{Shastry} with
equal strength of all exchange bonds. Again, the GS is magnetically
long-range ordered, see e.g. Refs.
\onlinecite{Mila,hartmann00,koga00a,lauchli02,rachid05}. However, the magnetic
LRO may be destroyed by a modification of bond strengths.
\cite{Shastry,Mila,hartmann00,koga00a,lauchli02,rachid05}           

The coordination number $z$ is quite large for the triangular ($z=6$) 
and SrCuBO ($z=5$ ) lattices, and that might be responsible for
the semi-classical GS LRO found for 
the HAF on these frustrated Archimedian lattices.
Thus, a ``non-magnetic'' quantum GS might be favored for lattices with lower
coordination number $z$.    
Indeed, a regular depletion of the triangular lattice by a factor
of 1/4 yields the  Archimedian kagom\'e lattice with coordination number
$z=4$. 
Contrary to the triangular lattice the GS of the 
spin-$\frac{1}{2}$ HAF 
on the kagom\'e lattice is most likely non-magnetic, see e.g. Refs.
\onlinecite{wir04,Lech,Wal,schmal_kag,singh07,white2010}.
Another frustrated model with low coordination number $z=3$ having a non-magnetic quantum GS 
is the HAF on the Archimedian
star lattice.\cite{wir04,star04,star07,star09}
Moreover, we mention that the HAF on the (non-Archimedian) square-kagom\'e
lattice with  $z=4$
has most likely also a non-magnetic quantum GS.\cite{Sidd,tomczak03,sp04,squago}
The above-mentioned depletion of the triangular lattice 
by a quarter is clearly not the only possibility.
As has been pointed out by D. Betts \cite{Betts} 
a regular depletion of the triangular lattice
by a factor of 1/7 yields another translationally invariant lattice, namely
the Archimedian maple-leaf lattice.\cite{Schmalfuss,wir04} 
The coordination number of this lattice is  $z=5$ and
lies between those of the triangular ($z=6$) and the 
kagom\'e ($z=4$) lattices. Moreover, there is a frustrated Archimedian
lattice with $z=4$, the so-called bounce lattice.\cite{wir04}
Both the  maple-leaf and the bounce lattices
might be candidates for non-magnetic GSs.
However, there are indications from previous studies (based on exact 
diagonalizations of finite-sized lattices) of semi-classical GS 
magnetic LRO \cite{Schmalfuss,wir04} in the spin-$\frac{1}{2}$ 
HAF on these lattices. 
This conclusion was drawn based on only two finite lattices of
$N=18$ and $N=36$ sites and therefore the conviction in the conclusions is
lessened. 

It is interesting  to point out that the discussion of the magnetic
properties of the HAF on Archimedian lattices is not only a 
challenging theoretical problem of quantum many-body physics
but that it is also strongly relevant to experiment. Indeed, 
most of these lattices are found to be underlying lattice
structures of the magnetic ions of various magnetic compounds, such as
CaV$_4$O$_9$\cite{Tan},
SrCu$_2$(BO$_3$)$_2$\cite{Kag}, or
[Fe$_3$($\mu_3$-O)($\mu$T-OAc)$_6$-(H$_2$O)$_3$][Fe$_3$($\mu_3$-O)($\mu$-OAc)$_{7.5}$]$_2 \cdot$7
H$_2$O.\cite{star}
Recently  there has been also synthesized 
the magnetic compound
$\mathrm{M_x}[\mathrm{Fe}(\mathrm{O}_2\mathrm{CCH}_2)_2\mathrm{NCH}_2\mathrm{PO}_3]_6\cdot 
\mathrm{n H}_2\mathrm{O}$ (Ref. \onlinecite{cave})
with maple-leaf lattice structure as well as hybrid cobalt hydroxide
materials\cite{price2011} with maple-leaf like lattice structure which may stimulate increasing interest in
this lattice.
Moreover,  in the natural mineral spangolite
(Cu$_6$Al(SO$_4$)(OH)$_{12}$Cl$\cdot$3H$_2$0)
the magnetic copper ions sit on the lattice sites of the maple-leaf
lattice.\cite{spango1,fennell}
In particular, spangolite is a very interesting magnetic system, since magnetic copper ions
carry spin $1/2$ and the experimental data indicate that strong fluctuations at
low temperatures are present which may prevent magnetic
ordering.\cite{fennell} In Ref. \onlinecite{fennell} Fennell et al.
propose that the spin-half Heisenberg model on the maple-leaf 
lattice with five different exchange integrals is the relevant model 
for this material.

\begin{figure}[ht]
\begin{center}
\epsfig{file=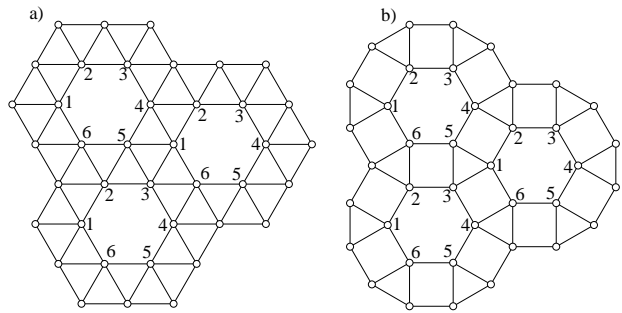,scale=0.35,angle=0.0}
\end{center}
\caption{Archimedian Lattices: (a) 1/7-depleted triangular (maple-leaf) lattice. (b) bounce lattice. 
The six equivalent sublattices for both lattices are indicated by `1', `2', `3', `4', `5', `6'} 
\label{fig1}
\end{figure}

\begin{figure}[ht]
\begin{center}
\epsfig{file=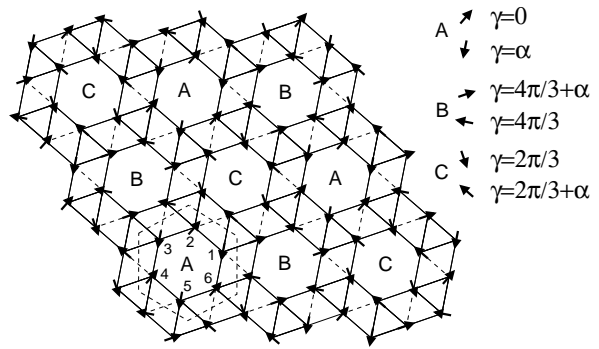,scale=0.25,angle=0.0}
\end{center}
\caption{Illustration of the classical GS of 
$J-J'$ model (Eq.~(\ref{ham}); $J$ -- solid lines; $J'$ -- dashed lines).
The numbers 1, 2, 3, 4, 5, and 6
denote the six sites in the geometrical unit cell. The magnetic unit cell
contains three  geometrical unit cells $A$, $B$ and $C$. 
}
\label{fig2}
\end{figure}

In this article we will discuss the GS properties of the spin-$\frac{1}{2}$
HAF on the maple-leaf and the bounce lattices, see Fig.\ref{fig1}. 
These lattices are related to each
other because the bounce lattice is equivalent to a bond-depleted maple-leaf
lattice, see Ref.~\onlinecite{wir04} and Fig.~\ref{fig2}.   
Therefore we will consider a generalized spin-$\frac{1}{2}$
$J$-$J'$ HAF
\begin{eqnarray}
\label{ham}
H&=&J\sum_{\langle ij \rangle}{\bf s}_{i} \cdot {\bf s}_{j}
+J' \sum_{[ ij ]}{\bf s}_{i} \cdot {\bf s}_{j} \; ; \; J>0 \; ; \; J' \ge 0
\;,
\end{eqnarray}
where ${\langle ij \rangle}$ runs over all nearest-neighbor (NN) bonds of the bounce lattice, 
and ${[ ij ]}$ runs over all additional NN bonds present in the maple-leaf
lattice, cf. Figs.~\ref{fig2} and \ref{fig3}.

A well-established method that can deal effectively with the GS properties of 
infinite 2D quantum magnets  is given by the coupled cluster method (CCM)
 (see, e.g., Refs.~\onlinecite{Bi:1998_b,zeng98,bishop00,farnell04,ijmp2007}
and references cited therein).
The accuracy and effectiveness of this method in relation to the investigation 
of frustrated quantum spin systems has been strongly improved by the
implementation of a parallelized CCM code \cite{cccm} that carries out
high-order CCM calculations. In particular,  quantum phase transitions in 2D
quantum spin systems that are driven by frustration
can be studied by this 
method.\cite{krueger00,krueger01,rachid05,Schm:2006,bishop08,bishop08a,rachid08,bishop09,richter10}\\

\section{The classical ground state}
\label{clas}
We start with a brief illustration  of the classical GS of the model (see
Fig.~\ref{fig2})
i.e. the ${\bf s}_i$ are treated as classical vectors of length $s$.
The classical GSs in the limits $J'=1$ and $J'=0$ have already 
been discussed in Ref.~\onlinecite{wir04}. Starting from the 
information on the classical GS provided there  it  can be found
easily  for the $J-J'$ model.  
The geometrical unit unit cell (hexagon) contains 6 sites labeled by the
running index $n=1,...,6$.
The magnetic unit cell is three times larger.
Within one geometrical unit cell the pitch angle $\alpha$ between neighboring spins
(e.g. spin on sites 1 and 2) around the hexagon is given
by $\alpha=
 \pi - \arctan(\frac{J'\sqrt{3} }{4J-J'})$.
Next-nearest neighbor spins on a hexagon (e.g. spins on sites 1 and 3) are
parallel. 
Equivalent spins in two neighboring unit cells are rotated by a fixed angle
$\pm 2\pi/3$, see the spin directions in unit cells $A$, $B$, $C$ shown in
Fig.~\ref{fig2}.
In the limits $J'=0$ (bounce lattice), $J'=1$ (maple-leaf lattice) or
$J'\to \infty$ one has $\alpha=\pi$ and $E=-3JNs^2/2$, $\;\alpha=5\pi/6$ and
$E=-(1+\sqrt{3})JNs^2/2$ or $\;\alpha \to \pi/3$ and
$E \to -J'Ns^2/2$, respectively.    \\

\section{The Methods}
\subsection{Coupled Cluster Method (CCM)}
We start with  a brief illustration of the main features of the CCM.  For a
general  overview on the CCM the interested reader is referred, e.g., to
Refs.~\onlinecite{bishop91a,zeng98,farnell04} and  for details of the CCM computational algorithm 
for quantum spin systems (with
spin quantum number $s=1/2$) to
Refs.~\onlinecite{zeng98,bishop00,krueger00,ijmp2007}.
The starting point for a CCM calculation
is the choice of a normalized model or reference state $|\Phi\rangle$,
together with a set of mutually commuting
multispin creation operators $C_I^+$ which are defined over a complete set of
many-body configurations $I$. The operators
$C_I$ are the multispin destruction operators
and are defined to be the Hermitian
adjoint of the $C_I^+$. We choose $\{|\Phi\rangle;C_I^+\}$ in such a way
that we have $\langle\Phi|C_I^+=0=C_I|\Phi\rangle$, $\forall I\neq 0$.
Note that the CCM formalism corresponds to the thermodynamic limit $N\rightarrow\infty$.

For spin systems, an appropriate choice for the CCM model state $|\Phi\rangle$
is often a classical spin state, in which the most general
situation is one in which each spin can point in an arbitrary direction.
We then perform a local coordinate transformation such that all spins are 
aligned in negative $z$-direction in the 
new coordinate frame.\cite{krueger00,krueger01,rachid05} As a result we have
\be 
   |\Phi\rangle=|\downarrow\downarrow\downarrow\cdots\rangle; \quad C_I^+=s_i^+,\,\,
s_i^+s_{j}^+,\,\, s_i^+s_{j}^+s_{k}^+,\cdots
,\ee
(where the indices $i,j,k,\dots\;$ denote arbitrary lattice sites) for the model state and the 
multispin creation operators which
now consist of spin-raising operators only.

The CCM parameterizations of the ket and bra ground
states are given by
\begin{eqnarray}
\label{ket}
H|\Psi\rangle = E|\Psi\rangle\,\, ; \qquad  \langle\tilde{\Psi}|H = E
\langle\tilde{\Psi}| \; \; ;\nonumber\\
|\Psi\rangle = e^S|\Phi\rangle\,\, ; \qquad S = \sum_{I \neq
0}{\cal S}_IC_I^+\; \; ; \nonumber\\
\langle\tilde{\Psi}| =  \langle\Phi|\tilde{S}e^{-S}\,\, ; 
\qquad \tilde{S} =
1 + \sum_{I \neq 0}\tilde{\cal S}_IC_I^- \;.
\end{eqnarray}
The correlation operators $S$ and $\tilde {S}$ contain the  correlation coefficients
${\cal S}_I$ and $\tilde {\cal S}_I$ that we must determine.
Using the Schr\"odinger equation, $H|\Psi\ra=E|\Psi\ra$, we can now write
the GS energy as $E=\la\Phi|e^{-S}He^S|\Phi\ra$ and 
the magnetic order parameter (sublattice magnetization) is given
by $ M_s = -\frac{1}{N} \sum_i^N \la\tilde\Psi|s_i^z|\Psi\ra$, where $s_i^z$ is
expressed in the transformed coordinate system. (Note that all magnetic
sublattices carry the same sublattice magnetization.)

To find the ket-state and bra-state correlation coefficients we
require that the expectation value $\bar H=\langle\tilde\Psi|H|\Psi\rangle$
is a minimum with respect to the
bra-state and ket-state correlation coefficients, such that 
the CCM ket- and bra-state equations are given by
\begin{eqnarray}
\label{ket_eq}
\langle\Phi|C_I^-e^{-S}He^S|\Phi\rangle = 0 \qquad \forall I\neq
0\\
\label{bra_eq}
\langle\Phi|\tilde{\cal S}e^{-S}[H, C_I^+]e^S|\Phi\rangle = 0 \qquad \forall
I\neq 0.
\end{eqnarray}
The problem of determining the CCM equations now becomes a
{\it pattern-matching exercise} of the $\{C_I^-\}$ to
the terms in $e^{-S}He^S$ in Eq.~(\ref{ket_eq}).
 
The CCM formalism is exact if we take into account all possible multispin
configurations in the correlation operators $S$ and $\tilde S$. This is, however, generally not 
possible for quantum many-body models including that studied here. 
We must therefore use the most common  
approximation scheme to truncate the expansion of $S$ and $\tilde S$ 
in the Eqs.~(\ref{ket_eq}) and (\ref{bra_eq}), namely the LSUB$n$ scheme, where we 
include only $n$ or fewer correlated spins in all configurations (or lattice animals in the language of graph theory) which 
span a range of no more than $n$ adjacent (contiguous) 
lattice  sites (for more details see Refs.
\onlinecite{bishop91a,bishop00,krueger00,ijmp2007}). For instance,
one includes multispin creation operators of one, two, three or four spins
distributed on arbitrary clusters of four contiguous lattice sites
for the LSUB4 approximation. The number of these fundamental configurations
can be reduced exploiting lattice symmetries. 
In the CCM-LSUB8 approximation we have finally $330369$ fundamental configurations.

Since the LSUB$n$ approximation becomes exact for $n \to \infty$, it is useful 
to extrapolate the `raw' LSUB$n$ data in the limit $n \to \infty$. 
An appropriate extrapolation rule for the order parameter of systems showing 
GS magnetic LRO is given by  
 $M_s(n)= a_0+a_1(1/n)+a_2(1/n^2)$ (see, e.g.,
Refs.~\onlinecite{bishop00,farnell04,ijmp2007})
 where the results  of the LSUB2,4,6,8 approximations are used for the 
extrapolation. For the GS energy per spin, 
$E(n)/N = b_0 + b_1(1/n^2) + b_2(1/n^4)$ is a well-tested extrapolation
ansatz.\cite{bishop00,krueger00,krueger01,farnell04,ijmp2007}\\

\subsection{Exact diagonalization (ED)} \label{ED}
The Lanczos ED is a well-established many-body method,
see, e.g., Ref.~\onlinecite{poilblanc}. Hence we can restrict our
discussion of the method to some specific  features relevant for our problem.
The  ED has been successfully applied to 2D frustrated quantum spin systems, see
e.g. Refs.~\onlinecite{wir04,Wal,schulz,ED40}.
Here we follow the lines of Ref.~\onlinecite{wir04}.
For the system under consideration there are only two appropriate finite lattices, namely one with $N=18$ (see,
Fig.~2 of Ref.~\onlinecite{Schmalfuss}) and
another one with $N=36$ (see Fig.~\ref{fig3}).
Note that these finite lattices do not have the the full lattice symmetry of
the infinite lattice. Moreover 
the unit cell of these lattices 
is fairly large, namely it contains six sites.
Hence, we consider the ED data as a complementary information to
confirm or to question the  CCM results corresponding to the
thermodynamic limit $N\rightarrow\infty$.
For the finite-size order parameter we use\cite{wir04}
\begin{equation} \label{mdef}
  m^+=\sqrt{\frac{(M^+)^2}{N^2}}
     =\left(\frac{1}{N^2}
     \sum_{i,j}^{N}  {}|\langle{\bf s}_{i}{\bf s}_{j}\rangle|\right)^{1/2}
     \, .
\end{equation}
We extrapolate the GS energy and the order parameter as described in
Ref.~\onlinecite{wir04} (see also Refs.~\onlinecite{schulz,ED40} and
references therein). 
However, the results of these ED extrapolations has to be taken with caution, since
they are based only on two data points.\\

\section{Results}
\begin{figure}[ht]
\begin{center}
\epsfig{file=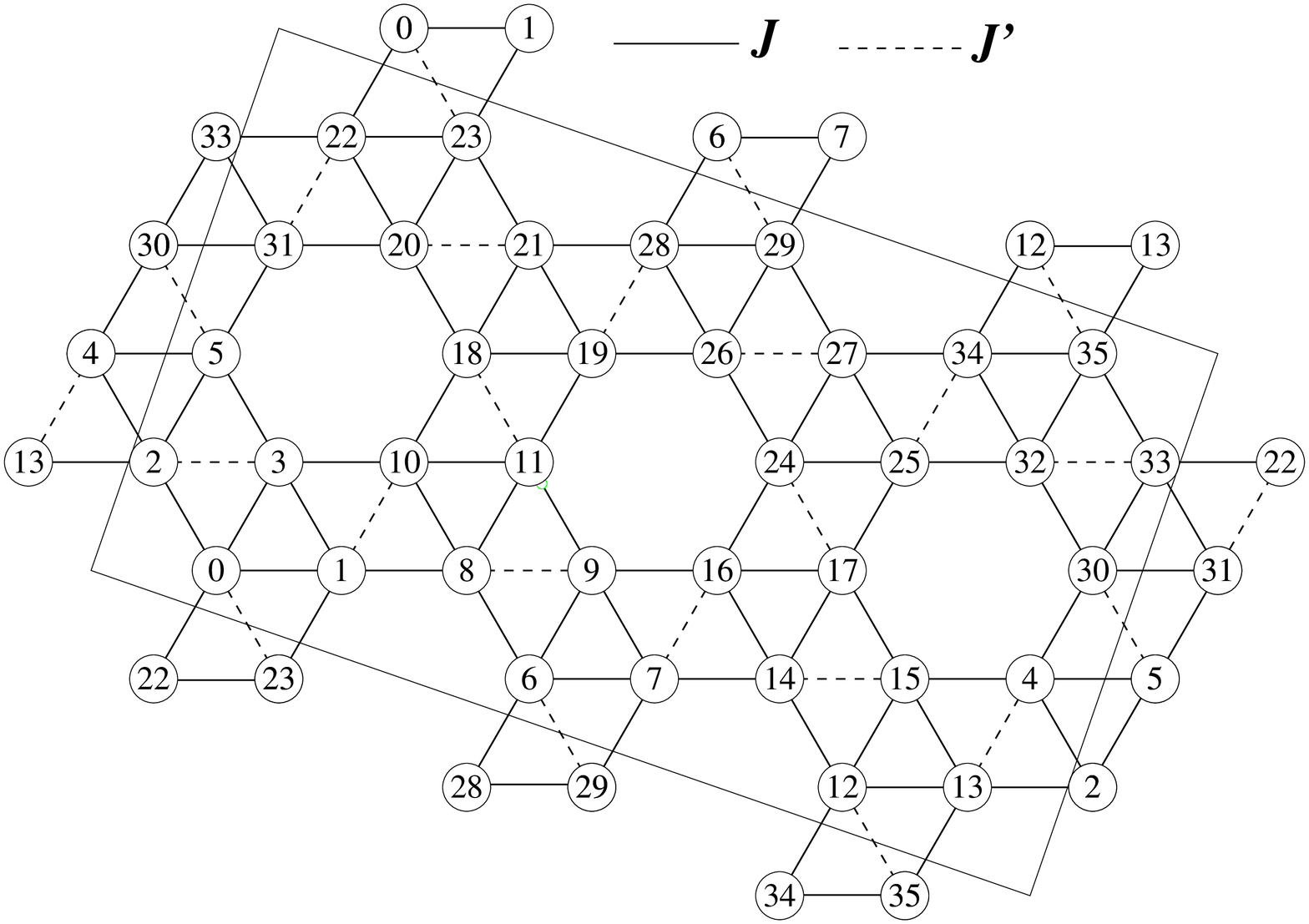,scale=0.31,angle=0.0}
\end{center}
\caption{The finite maple-leaf lattice of 36 sites imposing periodic boundary
conditions used for the ED calculations. Removing the dashed $J'$ bonds the
the exchange pattern corresponds to the bounce lattice of $N=36$ sites. 
}
\label{fig3}
\end{figure}

Henceforth, we set $J=1$ and we consider $J'$ as the active parameter 
in the model of Eq.~(\ref{ham}). We apply high-order CCM up to LSUB8 and these 
infinite-lattice results are complemented by ED results of $N=18$ and $N=36$
sites, see Fig.~\ref{fig3}. We choose the classical canted state
illustrated in Sec.~\ref{clas} to be the CCM reference or model state.
As quantum fluctuations may lead to a ``quantum'' pitch angle that is 
different from the classical case, we consider the pitch angle $\alpha$ in the 
reference state as a free parameter. We then determine the quantum 
pitch angle $\alpha_{qu}$ by minimizing $E_{{\rm LSUB}n}(\alpha)$ 
with respect to $\alpha$ in each order $n$. 

\begin{table}
\caption{CCM-results for the spin-$1/2$ HAFM on the bounce ($J'=0$) and on the
maple-leaf ($J'=1$) lattices.
$E/N$ is the GS energy per spin and $M_s$ is the sublattice magnetization. 
The LSUB$n$ results are extrapolated using $E(n)/N = b_0 + b_1(1/n^2) +
b_2(1/n^4)$ for the GS energy and
$ M_s(n)= a_0+a_1(1/n)+a_2(1/n^2)$ for the sublattice magnetization.
}
{\begin{tabular}{@{}|c|c|c|@{}}\hline 
{\bf bounce}                  &$E/N$          &$M_s$                \\\hline
LSUB2                        &-0.521631      &0.404343           \\\hline
LSUB4                        &-0.546866      &0.339357            \\\hline
LSUB6                        &-0.553763      &0.298249            \\\hline
LSUB8                        &-0.556998      &0.265252            \\\hline
{\bf Extrapolated CCM}   &{\bf  -0.5605 }         &{\bf 0.1657}              \\\hline
\hline
{\bf maple-leaf}                 &$E/N$          &$M_s$                \\\hline
LSUB2                        &-0.483470      &0.405622         \\\hline
LSUB4                        &-0.512309      &0.338483           \\\hline
LSUB6                        &-0.520378      &0.297499           \\\hline
LSUB8                        &-0.523861      &0.265768            \\\hline
{\bf Extrapolated CCM}   &{\bf-0.5279 }         &{\bf  0.1690 }              \\\hline
\hline
\end{tabular}}
\label{table1}
\end{table}

We start with the case of the perfect Archimedian bounce and maple-leaf
lattices, and so we set $J'=0$ and $J'=1$, respectively. 
Results for the GS energy and sublattice 
magnetization are given for both lattices in Table~\ref{table1}. GS energies
agree well with the previously reported data.\cite{wir04,Schmalfuss} 
Furthermore, we confirm the previous findings that the GS is magnetically ordered.
However, due to quantum fluctuations and frustration the sublattice magnetization is
drastically reduced. Using our extrapolated CCM data (see
Table~\ref{table1})
we find that the sublattice magnetizations  
are only $33\%$ of the classical value for the bounce lattice and  $34\%$
of the classical value for the maple-leaf lattice. However, these values are
still clearly above the ED estimates (which are $22\%$ for the maple-leaf
and $27\%$ for the bounce lattice) of Ref.~\onlinecite{wir04}.
We believe that the CCM data reported here 
are more reliable than the ED estimates because these ED results were 
extrapolated using only two data points ($N=18$ and $N=36$)\cite{misprint},
see also Sec.~\ref{ED}.
These rather small values of the order parameter, which are significantly 
below that for the triangular lattice,\cite{Ber1,chub94,Cap,farnell01,wir04,farnell09}
indicate that the GS magnetic LRO is fragile, and one can speculate that 
slight modifications of the model parameters might lead to non-magnetic 
quantum GSs. We remark again that a related experimental 
material called spangolite \cite{fennell} did not appear to 
shown magnetic LRO, and that this experimental result also spurs
us on to evaluate the more general model.

\begin{figure}[ht]
\begin{center}
\epsfig{file=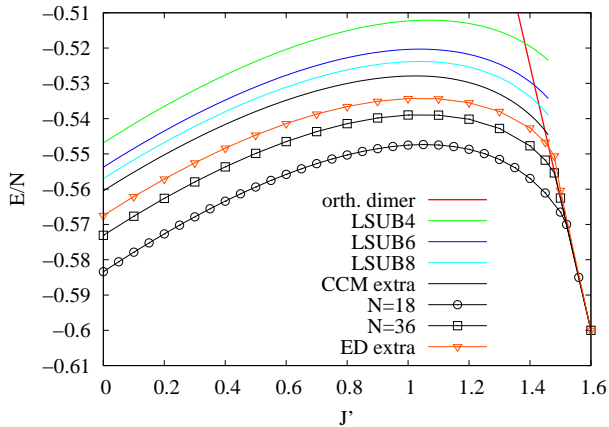,scale=0.66,angle=0.0}
\end{center}
\caption{CCM results for the GS energy per spin $E/N$ as function of $J'$.
The CCM extrapolated values in the limit $n \to \infty$
are obtained using the extrapolation scheme 
$E(n)/N = b_0 + b_1(1/n)^2 + b_2(1/n)^4$.
ED results for $N=18$ and $N=36$ as well as ED extrapolated values using the
extrapolation scheme
$E(N)/N = x_0 + x_1 N^{-3/2}$ are also shown.
The red straight line shows the exact energy of the orthogonal-dimer state.}
\label{fig4}
\end{figure}


\begin{figure}[ht]
\begin{center}
\epsfig{file=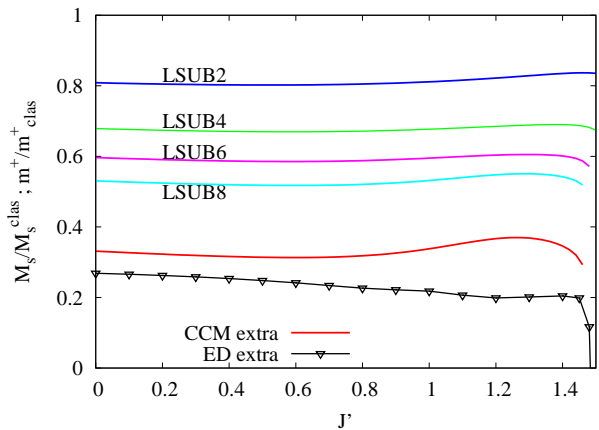,scale=0.66,angle=0.0}
\end{center}
\caption{CCM results for magnetic order parameter (sublattice magnetization)
$M_s$ as function of  $J'$. The data for the quantum model are scaled by its corresponding classical value
$M_s^{clas}=1/2$.
The extrapolated values in the limit $n \to \infty$ are found 
using the extrapolation scheme $ M_s(n)=c_0+c_1(1/n)+c_2(1/n)^{2}$. 
We also show 
extrapolated finite-lattice results for the ED order parameter $m^+$ 
using the
extrapolation scheme
$m^+(N) = y_0 + y_1 N^{-1/2}$.
}
  \label{fig5}
\end{figure}


The results for the GS energy per spin are shown in Fig.~\ref{fig4} as a 
function of $J'$. CCM LSUB$n$ results for the GS energy are 
clearly converging rapidly with increasing $n$ for all values of $J'$. 
We have also used our ED data for $N=18$ and $N=36$ to extrapolate them to $N
\to \infty$ (for the details of the extrapolation, see
Ref.~\onlinecite{wir04}). As already mentioned above
this ED extrapolation has to be taken with caution, since it is based only
on two data points.
We find a reasonable agreement between the CCM and ED data for the GS
energy.
As indicated by the ED data the GS energy becomes linear in $J'$ at
larger $J' \gtrsim 1.5$.  This is related to the existence of a dimer
product eigenstate where all $J'$ bonds carry a dimer singlet.\cite{misguich99}
We find that this singlet orthogonal-dimer 
eigenstate becomes the GS for  $J'>J'_c$.
Hence  our
model has much in common with the Shastry-Sutherland
model \cite{Shastry,Mila,hartmann00,koga00a,lauchli02,rachid05} 
that also demonstrates a similar exact orthogonal-dimer GS. 
We use  the intersection point between the extrapolated  CCM GS energy per
site
and the energy of the orthogonal-dimer
eigenstate given by $E_{OD}/N =-3J'/8$
to determine the transition point $J'_c=1.449$.
(Note that the corresponding value based on the ED data is
$J'_c=1.462$.)

Next we use the CCM results for the magnetic order parameter $M_s$ to discuss the
stability of magnetic LRO as a function of $J'$, see Fig.~\ref{fig5}. 
It is obvious  that the magnetic 
LRO persists in the whole region $0 \le J' \le J'_c$. 
This conclusion is supported by the extrapolated ED order parameter also
shown for comparison in Fig.~\ref{fig5}. 
Interestingly the dependence of the order parameter on $J'$ is fairly weak
over the whole region $0 \le J'\le J'_c$. Thus, the extrapolated CCM order
parameter varies only between  $29\%$ and $37\%$ of its classical value
$M_s^{clas}=1/2$.
This behavior might be interpreted as balanced interplay between
increasing of frustration and increasing of the number of nearest neighbors when
$J'$ is growing.      
Our data for the order parameter lead to the conclusion 
that there is a direct first-order transition to the 
 magnetically disordered orthogonal-dimer singlet GS.

\begin{figure}[!ht]
\begin{center}
\epsfig{file=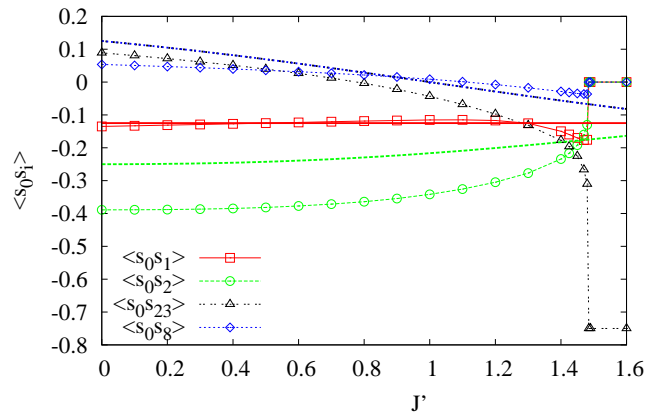,scale=0.7,angle=0.0}
\end{center}
\caption{ED results for some selected spin-spin correlation functions 
$\langle {\bf s}_0 {\bf s}_i\rangle $, $i=1,2,8,23$, 
for the finite lattice of $N=36$ sites shown in
Fig.~\ref{fig3}. The results for the quantum $s=1/2$ model are given by thin lines with
symbols. The corresponding classical results are given by thick lines with
the same color  without symbols. The location of sites '0' and 'i' in
$\langle {\bf s}_0 {\bf s}_i\rangle$ can be found in Fig.~\ref{fig3}.
Note that the classical curves for $\langle {\bf s}_0 {\bf s}_8\rangle$ and
$\langle {\bf s}_0 {\bf s}_{23}\rangle$ coincide. 
}
\label{fig6}
\end{figure}
\begin{figure}[!ht]
\begin{center}
\epsfig{file=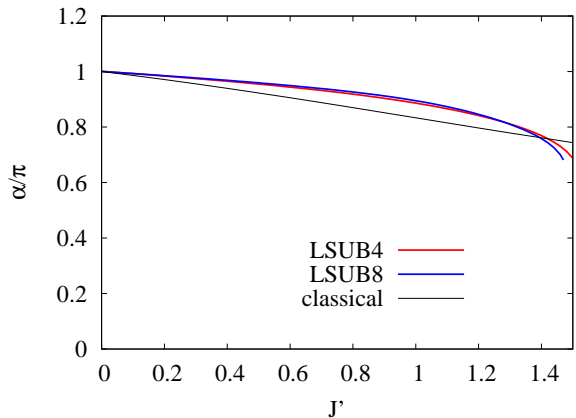,scale=0.65,angle=0.0}
\end{center}
\caption{Results for the quantum pitch angle $\alpha_{qu}$  
as a function of $J'$ calculated within  CCM-LSUB$n$ approximation 
with $ n=4$ and $8$. The classical result $\alpha_{cl}$ is shown for 
the sake of comparison.}
\label{fix4}
\end{figure}

An additional confirmation of the above discussed behavior comes from the ED
data for 
the spin-spin correlation functions $\langle {\bf s}_i {\bf s}_j\rangle $ presented in
Fig.~\ref{fig6}.
Again we see that the variation of the correlation functions with $J'$ is weak
almost
up to the transition point $J'_c$. Moreover,  for $J' < J'_c$ the correlation functions of
the quantum model behave similar to those of the classical model.

Finally, in Fig.~\ref{fix4} we compare results for the classical pitch angle $\alpha_{cl}$, see
Sec.~\ref{clas}, and for the quantum pitch angle  $\alpha_{qu}$ calculated
by the CCM . 
We see that both $\alpha_{cl}$ and $\alpha_{qu}$ are
close to each other and that there is only a slight variation of the pitch
angle with $J'$ (the value of the quantum pitch
angle in CCM-LSUB8 approximation is $\alpha_{qu}=\pi$ at $J'=0$,
$\alpha_{qu}\approx 0.895 \pi$
and it is still $\alpha_{qu} \approx 0.714\pi$ at
$J'=J'_c$).
Moreover, the  LSUB4 and LSUB8 data for  $\alpha_{qu}$
almost coincide. 

Altogether,
our data for the order parameter, the spin-spin correlation functions, and the
pitch angle lead to the conclusion 
that there is most likely a direct first-order transition from a magnetically
ordered state with spin orientations similar to those of the classical GS to
the  
magnetically disordered orthogonal-dimer singlet GS.

\begin{figure}[ht]
\begin{center}
{\epsfig{file=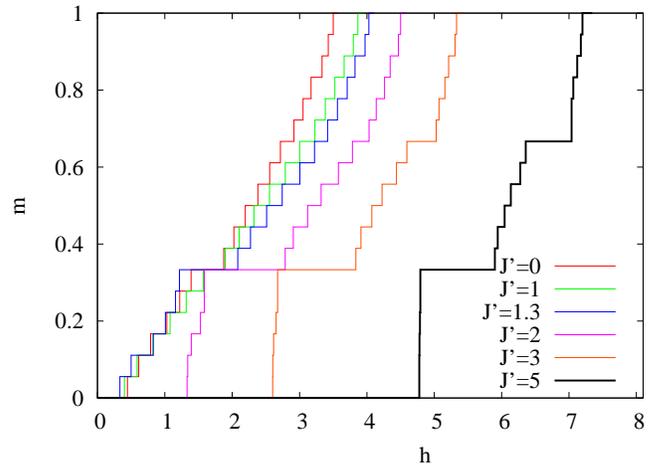,width=3.6in,angle=0}}
\end{center}
\caption{Results for the relative magnetization $m=M/M_{sat}$ versus applied field 
of strength $h$ obtained via ED for $N=36$, where $M_{sat}$ is the saturation
magnetization.}
  \label{fig8}
\end{figure}

\begin{figure}[ht]
\begin{center}
{\epsfig{file=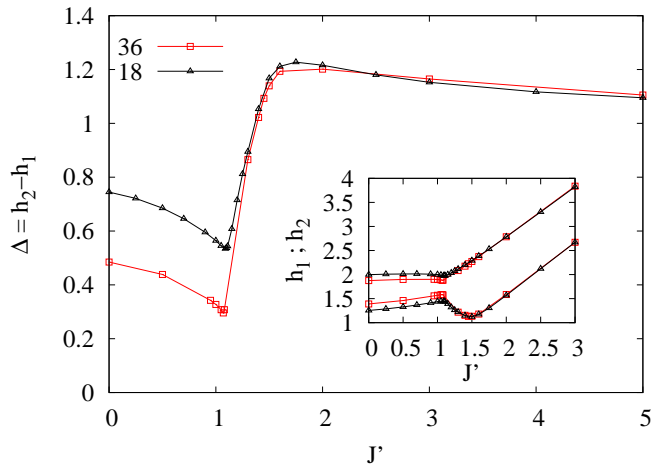,width=3.6in,angle=0}}
\end{center}
\caption{ED results ($N=18,36$) for the $m$=1/3 plateau. Main panel:
Width $\Delta=h_2-h_1$  of the plateau versus $J'$. Inset:
End points $h_1$ and $h_2$ of the 
plateau versus $J'$.}
  \label{fig9}
\end{figure}

\begin{figure}[ht]
\begin{center}
{\epsfig{file=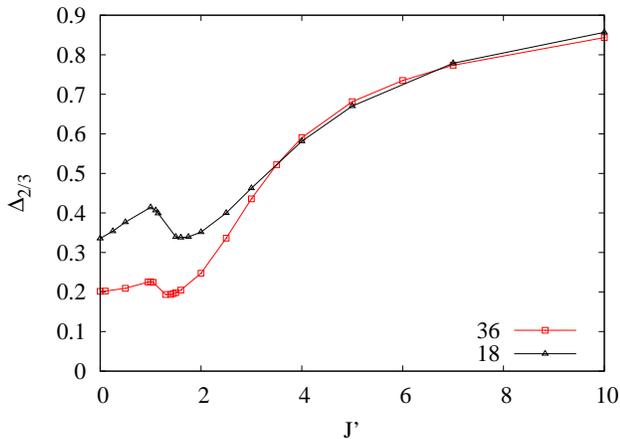,width=3.4in,angle=0}}
\end{center}
\caption{ED results ($N=18,36$) for the width  of the m=2/3 plateau in
dependence on $J'$.}
  \label{fig10}
\end{figure}

\section{Magnetization process}
The magnetization process of frustrated quantum magnets has attracted much
attention due to the discovery of exotic parts of the magnetization curve, such as
plateaus and
jumps, see e.g. Refs.~\onlinecite{nishi,Hon1999,LhuiMi,jump,HSR04}.
The magnetization curves for the pure bounce ($J'=0$) and maple-leaf 
($J'=1$) 
HAF were discussed already in Ref.~\onlinecite{wir04} based on ED data for
$N=36$, where no indications for plateaus and jumps were found.
On the other hand,
we have already seen that the interpolating maple-leaf/bounce
lattice AF model  considered here for larger values of $J'$
has much in common with the Shastry-Sutherland
model. In particular, that both have at zero field a orthogonal-dimer
singlet ground state. 
It is well  known
that the magnetization curve of the  material SrCu$_2$(BO$_3$)$_2$
as well as that of the corresponding Shastry-Sutherland model possesses a series of
plateaus, see e.g.
Refs.~\onlinecite{wir04,Kage,kodama,misguich,infinite,mila}.
Motivated by this we study in this section the  magnetization curve
$M(h)$ (where $M$ is the total
magnetization  and  $h$ is
the strength of the external magnetic field)
for the interpolating maple-leaf/bounce lattice AF model 
using ED for $N=18$ and $N=36$ sites. 
ED results  for the relative magnetization $m=M/M_{sat}$ versus magnetic
field $h$ for  
$N=36$ sites are shown in Fig.~\ref{fig8}. 
In accordance with previous results\cite{wir04} we do not see indications
for a plateau for $0 < J' \le 1$.
Moreover it is obvious, that the finite-size singlet-triplet gap determining the
size of the first
plateau at $m=0$ is small at $J'=0$, $J'=1$ and $J'=1.3$.  That
corresponds to our finding of a magnetically ordered GS for these values of
$J'$, and therefore the $m=0$ plateau should disappear for $N \to \infty$. 
However, a finite $m=0$ plateau exists 
in that parameter region where the orthogonal-dimer singlet state is the
zero-field GS, since this GS is gapped.

Similar as for the Shastry-Sutherland model there is a well-pronounced 1/3 plateau
appearing at larger values of $J'$. 
Interestingly, this plateau emerges already for values of $J'$ below
$J'_c \approx 1.5$.
In Fig.~\ref{fig9} we show the width of the 1/3 plateau versus $J'$ (main
panel) as well as the end points of the plateau (inset) for
$N=18$ and $N=36$.
We observe a significant change at $J' = J'_{p_{1/3}}\approx 1.07$. 
Below $ J'_{p_{1/3}}$ the typical finite-size behavior appears, i.e. the plateau
width shrinks with system size $N$ and it vanishes at $N \to \infty$.
By contrast, for $J'> J'_{p_{1/3}}$ there are almost no finite-size effects.
The 
plateau width increases rapidly up to about $J' \approx J'_c$ and then it remains
almost constant for $J' \gtrsim J'_c$, where the end points of the plateau
grow linearly with $J'$.
Moreover, from Fig.~\ref{fig8} we find that 
for larger values of $J' \gtrsim 2$ there is a jump-like transition
between the $m=0$ and the $m=1/3$ plateaus.

As already mentioned above, for the Shastry-Sutherland model a series of
plateaus
was observed.
For the model under consideration here we find indications for  a
second plateau at $m=2/3$, see Figs.~\ref{fig8} and \ref{fig10}. This plateau emerges only for quite large values
of $J' > J'_{p_{2/3}}\approx 3$. Again we observe very weak finite-size effects of the
plateau width for $J' > J'_{p_{2/3}}$ (Fig.~\ref{fig10}). Furthermore, our ED data suggest 
an almost direct jump from the  $m=2/3$ plateau to saturation $m=1$.

Finally we have to mention, that our finite-size analysis of the plateaus
naturally could miss other plateaus present in infinite systems, see e.g.
the discussion of the ED data of the $m(h)$ curve of the Shastry-Sutherland
model in Ref.~\onlinecite{wir04}. Hence, the study of the magnetization process
needs further attention based on alternative methods.\\

\section{Conclusions}
In this article we have treated a $J$-$J'$ spin-half  HAF interpolating
between the HAF on the maple-leaf ($J'=J$)  and the bounce
lattice ($J'=0$). Moreover, we also discuss the GS for larger values $J'>J$.  
This antiferromagnetic system is geometrically
frustrated and it is related to several magnetic materials which have been
experimentally investigated recently.\cite{cave,price2011,fennell}
On the classical level the ground state is
a commensurate non-collinear antiferromagnetic state.
To study the quantum GS of the  spin-half model     
we  use the  CCM for infinite lattices and the ED for finite lattices of $N=18$ and $N=36$ sites.

We find evidence for a semi-classical magnetically ordered commensurate non-collinear GS
in a  wide range of the exchange ratio $0 \le J'/J \le J'_c/J
\approx 1.5$.
However, due to frustration and quantum fluctuations the sublattice
magnetization amounts about $30\%$ of the classical value only.
Importantly, we find that at $J'_c \approx 1.45 J$ there is a (most likely) first-order
transition to a magnetically disordered orthogonal-dimer singlet product GS
which is the exact GS  for $J'/J> J'_c/J$. Therefore, the considered  model is somewhat similar to
the Shastry-Sutherland model. This
similarity is also observed in the 
magnetization curve.
Based on ED data we find evidence for plateaus at zero magnetization for $J'/J> J'_c/J$
(which is related to the gapped orthogonal-dimer singlet GS), at  $1/3$ of saturation for  $J'
\gtrsim 1.07 J$ and at  $2/3$ of saturation for 
$J' \gtrsim 3 J$. 
The transition to the $1/3$ plateau and from the $2/3$ plateau to saturation
can be jump-like.      
Further plateaus not compatible to the system sizes $N=18$ and $N=36$ and
therefore missed in the ED study
may appear in the infinite system.

Our results may also stimulate other studies on this interesting 2D frustrated
quantum system using  
other approximate methods to compare and contrast
our results.
In particular, the influence of modifications of the exchange bonds which
might be relevant for real-life materials may lead to a destabilization of
magnetic order. Knowing that the corresponding Shastry-Sutherland model
exhibits a series of magnetization plateaus  the search for further plateaus
may be  also an interesting problem to be studied by different means.\\

{\bf Acknowledgment}
For the exact digonalization J. Schulenburg's {\it  spinpack} was used.
The authors are indebted to T. Fennell for providing information on
spangolite prior to publication.


\begin{thebibliography}{99}
\bibitem{misprint}
Note here that there is a misprint in Ref.~\onlinecite{wir04} concerning the
extrapolated GS energy per bond of the mapel-leaf lattice. The correct value reads
$E_0/bond=0.2137$.
\end{thebibliography}
\end{document}